# Decoherence Estimation of Superconducting Qubit

Yoav Koral[1], Shilo Avraham[2], Manimuthu Periyasamy[1,3], Shmuel E. Schacham[1], Eliyahu Farber[1,3]

## Abstract

Decoherence of quantum bits arises primarily from the parasitic resistance within the qubit. This study presents the analysis of the decoherence process due to physical interactions between the qubit photons and parasitic resistance atoms, utilizing exclusively the Caldeira-Leggett electrical model, without relying on external Hamiltonians. The analysis shows a good agreement between the model of the electrical noise and the Johnson-Nyquist noise. The emission and absorption rates of the qubit's coherence loss, required for the Lindblad master equation that approximates the decoherence, are obtained. A numerical substitution in the analysis result yields a strong correlation with previous measurements. The present analysis enables also the derivation of the appropriate circuit characteristics for future simulations.

## 1. Introduction

The fundamental unit of information in quantum computing is the quantum bit, commonly referred to as a qubit. The most common implementation of qubits is by superconducting Josephson Junctions. These qubits are controlled using an external voltage source [1]. Once a qubit is set to a specific quantum state, a combination of the first two eigenstates, $|0\rangle$ and $|1\rangle$, it starts to experience decoherence [2], the process where the coherence between the two eigenstates is lost. This loss of coherence is primarily due to the parasitic resistance within the qubit [3], which has two main causes: the emission of photons into the resistor with the involvement of atoms in this process, as well as the absorption of photons that are excited by the thermal vibrations of the resistor's atoms.

The decoherence process has been investigated based on several theoretical approaches. Breuer and Petruccione [4] give reference to prior works [5, 6] that utilize a Hamiltonian based on a single qubit interacting with its environment [7], and the Caldeira-Leggett model [8], known as the spin-boson model Hamiltonian. Alternatively, the Ising model was used to describe the decoherence process [9]. Recent works [10, 11, 12] investigate decoherence by pure mathematical analysis based on physical and mathematical models which have no correlation with any electrical circuit model. The first to use an electric field model to represent the internal resistance of the qubit, to derive quantum fluctuations, was Devoret [13]. He employed the Foster model [14] to replace the resistance of the qubit by an infinite number of parallel LC resonators connected in series. A subsequent effort by Cataneo and Paraoanu [15] provided a comprehensive review of the decoherence process, for both a single qubit and two qubits connected to external resistors, using small serial capacitors. However, this work analyzed only the influence of external resistors on qubit's decoherence, ignoring the part of qubit's internal resistance.

Along the lines of the previous works, mainly that of Cataneo and Paraoanu [15], we introduce a comprehensive analysis of the Hamiltonian for a qubit with parasitic resistance, based on the Foster model of the resistance. We calculate the parameters of the model and employ the Lindblad master equation [16] to analyze and simulate the decoherence process.

## 2. Calculation Process

To simplify the analysis of the decoherence process, it is necessary to assume that the qubit is in an initial quantum state, disconnected from the surroundings, as shown in Fig. 1.

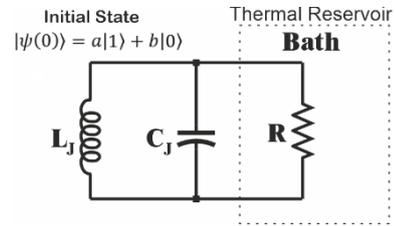

Fig. 1 • Simplified qubit model with internal losses/resistance

Resistors cannot be included in the circuit Hamiltonian due to lack of energy conservation. To overcome this problem, Devoret [13] replaced the resistors by an infinite number of LC resonators connected in series as shown in Fig. 2. The resonators differ from one another by their resonance frequencies, $1/\sqrt{L_k C_k}$, that equals $k\Delta\omega$ for the k'th resonator, and by the section admittance, $\sqrt{C_k/L_k}$ that its value is set using the required bath resistance function $R_B(\omega_k)$.

[1] SMQS Group, Department of Electrical & Electronic Engineering, Ariel University, ISRAEL
[2] School of Physics and Astronomy, Tel-Aviv University, ISRAEL
[3] Department of Physics, Ariel University, ISRAEL

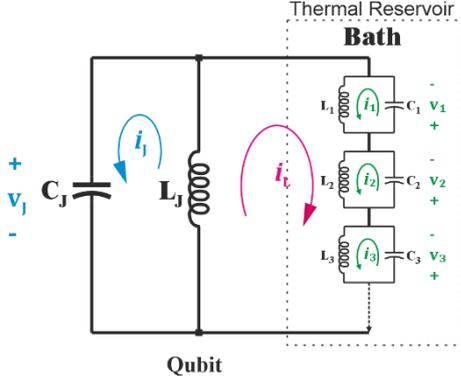

Fig. 2 • Qubit electrical model employing a thermal reservoir, representing the resistor by a set of resonators.

During transient time, following an input signal to the bath circuit with a finite number of resonators, the circuit behavior is similar to that of a resistor. With increasing time, the bath circuit behavior changes to a reactive network. As the number of the resonators increases to infinity, the transient time at which the bath circuit behaves as resistor, approaches to infinity as well. Therefore, the behavior of the network can be seen as an impedance with a significant resistive component and a negligible reactive component, but only within a limited frequency range, as discussed in section 4.

The calculation procedure for determining the decoherence process of the circuit shown in Fig. 2 is performed using six major steps:

1) System and bath Hamiltonian
2) Resistor's bath reactive model
3) Thermal noise of the resistor
4) Emission and absorption rates
5) Coherent wave decay constant
6) Solution of the system's Lindblad equation

These steps are discussed in the following sections.

## 3. System and Bath Hamiltonian

Using the Kirchhoff voltage equation for the loops, we obtain the following set of equations

$$L_J \frac{d}{dt}(i_J + i_L) + \frac{q_J}{C_J} = 0$$

$$L_k \frac{d}{dt}(i_k + i_L) + \frac{q_k}{C_k} = 0 \quad ; \quad k = 1,2 \ldots \infty \quad (1)$$

$$L_J \frac{d}{dt}(i_J + i_L) + \sum_{k=1}^{\infty} L_k \frac{d}{dt}(i_k + i_L) = 0$$

which leads to

$$L_J \frac{d^2 q_J}{dt^2} + L_J \frac{d^2 q_L}{dt^2} + \frac{q_J}{C_J} = 0$$

$$L_k \frac{d^2 q_k}{dt^2} + L_k \frac{d^2 q_L}{dt^2} + \frac{q_k}{C_k} = 0 \quad ; \quad k = 1,2 \ldots \infty \quad (2)$$

$$L_J \left( \frac{d^2 q_J}{dt^2} + \frac{d^2 q_L}{dt^2} \right) + \sum_{k=1}^{\infty} L_k \left( \frac{d^2 q_k}{dt^2} + \frac{d^2 q_L}{dt^2} \right) = 0$$

where

$$q_J = \int_{-\infty}^{t} i_J dt' \quad ; \quad q_k = \int_{-\infty}^{t} i_k dt' \quad ; \quad q_L = \int_{-\infty}^{t} i_L dt' \quad (3)$$

Using the relation

$$\begin{bmatrix} \dot{q}_1 \\ \dot{q}_2 \\ \dot{q}_3 \\ \vdots \\ \dot{q}_L \end{bmatrix} = \begin{bmatrix} L_J & 0 & 0 & 0 & 0 \\ 0 & L_1 & 0 & 0 & 0 \\ 0 & 0 & L_2 & 0 & 0 \\ 0 & 0 & 0 & \ddots & 0 \\ 0 & 0 & 0 & 0 & L_L \end{bmatrix}^{-1} \begin{bmatrix} \phi_1 \\ \phi_2 \\ \phi_3 \\ \vdots \\ \phi_L \end{bmatrix} \quad (4)$$

$$L_L = L_J + \sum_{k=1}^{\infty} L_k$$

and by combining the Euler-Lagrange equation and the Legendre transformation [17, 18], the circuit Hamiltonian can be derived as

$$\widetilde{H} = \widetilde{H}_S + \widetilde{H}_B + \widetilde{H}_I \quad (a)$$

$$\widetilde{H}_S = \frac{\Phi_J^2}{2L_J} + \frac{Q_J^2}{2C_J} \quad (b)$$

$$\widetilde{H}_B = \sum_{k=1}^{\infty} \left( \frac{\Phi_k^2}{2L_k} + \frac{Q_k^2}{2C_k} \right) \quad (c) \quad (5)$$

$$\widetilde{H}_I = \frac{\Phi_L^2}{2L_L} \quad (d)$$

The last term in equation (5) is the interaction Hamiltonian. It represents the energy in the $i_L$ current loop. This loop expresses the interaction between the qubit and the resonators in the bath. The absence of a capacitor in the interaction loop poses a challenge in describing the system by known operators and eigen-functions. Therefore, it is necessary to characterize the interaction flux as a dependent function of the fluxes of the other inductors in the circuit. A short way to do this, is to retreat to the equations set (2) and to calculate its Laplace transform in the following manner

$$s^2 L_J \bar{q}_J + s^2 L_J \bar{q}_L + \frac{\bar{q}_J}{C_J} = 0$$

$$s^2 L_k \bar{q}_J + s^2 L_k \bar{q}_L + \frac{\bar{q}_k}{C_k} = 0 \quad ; \quad k = 1,2 \ldots \infty \quad (6)$$

$$s^2 L_J (\bar{q}_J + \bar{q}_L) + s^2 \sum_{k=1}^{\infty} L_k (\bar{q}_k + \bar{q}_L) = 0$$

According to equation (4) and the third equation in (6)

$$\Phi_L = -\left( \Phi_J + \sum_{k=1}^{\infty} \Phi_k \right) \quad (7)$$

By substituting equation (7) into (5) and rearranging the Hamiltonian, the following expressions are obtained

$$\widetilde{H} = \widetilde{H}_S + \widetilde{H}_B + \widetilde{H}_I \qquad (a)$$

$$\widetilde{H}_S = \frac{\Phi_J^2}{2L_{AJ}} + \frac{Q_J^2}{2C_J} \qquad (b)$$

$$\widetilde{H}_B = \sum_{k=1}^{\infty}\left(\frac{\Phi_k^2}{2L_{Ak}} + \frac{Q_k^2}{2C_k}\right) \qquad (c) \qquad (8)$$

$$\widetilde{H}_I = \frac{\Phi_J}{L_L}\sum_{k=1}^{\infty}\Phi_k \qquad (d)$$

where

$$\frac{1}{L_{AJ}} = \frac{1}{L_J} + \frac{1}{L_L} \quad ; \quad \frac{1}{L_{Ak}} = \frac{1}{L_k} + \frac{1}{L_L} \qquad (9)$$

Using the definitions for the annihilation and creation operators [19], we obtain

$$\widehat{\Phi}_i = \frac{\phi_i}{\sigma_{\phi i}} \quad ; \quad \sigma_{\phi i} = \sqrt{\hbar\sqrt{L_i/C_i}}$$

$$\widehat{Q}_i = \frac{Q_i}{\sigma_{qi}} \quad ; \quad \sigma_{qi} = \sqrt{\hbar\sqrt{C_i/L_i}}$$

$$\boldsymbol{a} = \frac{1}{\sqrt{2}}(\widehat{\Phi} + i\widehat{Q}) \quad ; \quad \boldsymbol{a}^\dagger = \frac{1}{\sqrt{2}}(\widehat{\Phi} - i\widehat{Q}) \qquad (10)$$

$$i = AJ \text{ or } Ak$$

Subsequently, we proceed with the definition of the extended annihilation operators

$$\widetilde{\boldsymbol{a}} = \boldsymbol{a}\otimes_{k=N}^{1}\mathbb{I} \quad ; \quad \widetilde{\boldsymbol{b}}_{\boldsymbol{k}} = \mathbb{I}\otimes_{p=N}^{k+1}\mathbb{I}\otimes\boldsymbol{a}\otimes_{p=k-1}^{1}\mathbb{I} \quad (11)$$

where the indices decrease. Substituting equations (9) through (11) into equation (8), we obtain the Hamiltonian normalized to $\hbar$

$$\widetilde{H} = \widetilde{H}_S + \widetilde{H}_B + \widetilde{H}_I \qquad (a)$$

$$\widetilde{H}_S = \omega_{AJ}\left(\widetilde{\boldsymbol{a}}^\dagger\widetilde{\boldsymbol{a}} + \frac{1}{2}\right) \qquad (b)$$

$$\widetilde{H}_B = \sum_{k=1}^{\infty}\omega_{Ak}\left(\widetilde{\boldsymbol{b}}_{\boldsymbol{k}}^\dagger\widetilde{\boldsymbol{b}}_{\boldsymbol{k}} + \frac{1}{2}\right) \qquad (c) \qquad (12)$$

$$\widetilde{H}_I = \frac{1}{2}(\widetilde{\boldsymbol{a}}^\dagger + \widetilde{\boldsymbol{a}})\sum_{k=1}^{\infty}\omega_{Rk}(\widetilde{\boldsymbol{b}}_{\boldsymbol{k}}^\dagger + \widetilde{\boldsymbol{b}}_{\boldsymbol{k}}) \qquad (d)$$

where the frequencies are given by

$$\omega_{AJ} = 1/\sqrt{L_{AJ}C_J} \quad ; \quad \sigma_J = \sqrt{\hbar\omega_{AJ}L_J}$$

$$\omega_{Ak} = 1/\sqrt{L_{Ak}C_k} \quad ; \quad \sigma_k = \sqrt{\hbar\omega_{Ak}L_k} \qquad (13)$$

$$\omega_{Rk} = \frac{\sqrt{L_{AJ}L_{Ak}}}{L_L}\sqrt{\omega_{AJ}\omega_{Ak}}$$

Shifting the energy zero level of (12) by $\omega_{AJ} + \frac{1}{2}\Sigma_k\omega_{Ak}$ and reducing the annihilation operator $\boldsymbol{a}$'th matrix dimensions to 2x2, we obtain

$$\widetilde{H} = \widetilde{H}_S + \widetilde{H}_B + \widetilde{H}_I \qquad (a)$$

$$\widetilde{H}_S = -\frac{\omega_{AJ}}{2}\widetilde{\boldsymbol{\sigma}}_z \qquad (b)$$

$$\widetilde{H}_B = \sum_{k=1}^{\infty}\omega_{Ak}\widetilde{\boldsymbol{b}}_{\boldsymbol{k}}^\dagger\widetilde{\boldsymbol{b}}_{\boldsymbol{k}} \qquad (c) \qquad (14)$$

$$\widetilde{H}_I = \frac{1}{2}\widetilde{\boldsymbol{\sigma}}_x\sum_{k=1}^{\infty}\omega_{Rk}(\widetilde{\boldsymbol{b}}_{\boldsymbol{k}}^\dagger + \widetilde{\boldsymbol{b}}_{\boldsymbol{k}}) \qquad (d)$$

where $\widetilde{\boldsymbol{\sigma}}_x$ and $\widetilde{\boldsymbol{\sigma}}_z$ are the extended Pauli matrices [20].

## 4. Resistor's Bath Reactive Model

We employ the Fermi Golden Rule [21] for the calculation of emission and absorption rates. To that end, the energy states density associated with the qubit states are required. We start with the calculation of the bath frequency response. At every junction in the bath, the current is given by

$$i_B = C_k\frac{dv_k}{dt} + i_k \quad ; \quad k = 1,2\ldots\infty \qquad (15)$$

Hence, the Laplace transform of the total bath impedance is

$$Z_R = \frac{V_B}{I_B} = \sum_{k=1}^{\infty}\frac{1}{C_k}\frac{s}{s^2 + \omega_k^2} \quad ; \quad \omega_k = \frac{1}{\sqrt{L_kC_k}} \qquad (16)$$

This impedance represents the reactance part only. However, if the inductance and capacitance components of the resonator tend to zero and the number of resonators tends to infinity, the resistive component becomes significant. The ideal impedance of the bath network is equal to the qubit's parasitic resistance for all frequencies. However, it is not achievable with a reactive network. The practical alternative is to form a target impedance function that presents the required impedance in a band limited frequency response. This target impedance function has a real and an imaginary part as follows

$$Z_B(\omega) = R_B(\omega) + jX_B(\omega) \qquad (17)$$

To create the target resistance, we define a set of parallel resonators, where the resonance frequency of the k'th resonator equals to $k\Delta\omega$, and its section admittance, $\sqrt{C_k/L_k}$, is set so that the section bandwidth will be $2\Delta\omega$. In that way we form the required bath resistance $R_B(\omega_k)$:

$$Q_k = R_B(\omega_k)\sqrt{\frac{C_k}{L_k}} = \frac{\pi}{2}k \quad ; \quad \omega_k = k\Delta\omega \qquad (18)$$

where $Q_k$ is the section's quality factor. The scattering parameter $S_{21}$ of the first five resonators of the bath circuit with $k = 1,\ldots,5$, as a function of the normalized frequency $\omega/\omega_1$, are shown in Fig. 3. Each resonator is separately connected as a terminated 2 port circuit with a $R_B(\omega_k)$ resistor.

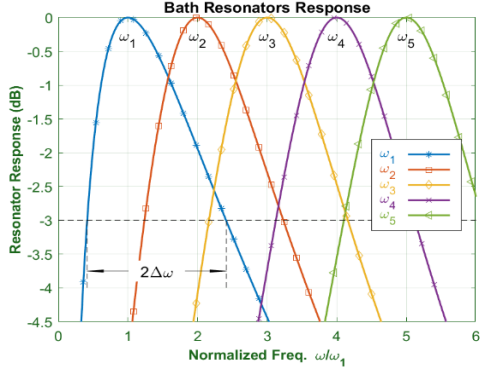

Fig. 3 • Frequency response $S_{21}$ of first 5 resonators of bath circuit vs. normalized frequency $\omega/\omega_1$

Using (16) and (18), we derive the capacitance and inductance of the resonator's components :

$$C_k = \frac{\pi}{2\,\Delta\omega\,R_B(\omega_k)} \quad ; \quad L_k = \frac{2\,R_B(\omega_k)\Delta\omega}{\pi\,\omega_k^2} \quad (19)$$

Substituting $C_k$ into $Z_R$ in equation (16), increasing the number of resonators to infinity, we get

$$Z_R(\omega) = \frac{-j}{\pi}\mathcal{P}\int_{-\infty}^{\infty}\frac{R_B(\omega_k)}{\omega - \omega_k}d\omega_k = i\,X_B(\omega) \quad (20)$$

where $\mathcal{P}$ denotes the Cauchy principal value integral [22]. The right-hand expression is derived from the Kramers-Krönig relations [23], which are applicable for every impedance that can be implemented and contains both real and imaginary parts

$$R_B(\omega) = \frac{1}{\pi}\mathcal{P}\int_{-\infty}^{\infty}\frac{X_B(\omega')}{\omega' - \omega}d\omega' \quad (a)$$

$$X_B(\omega) = -\frac{1}{\pi}\mathcal{P}\int_{-\infty}^{\infty}\frac{R_B(\omega')}{\omega' - \omega}d\omega' \quad (b) \quad (21)$$

Thus, using equation (21)a, we can conclude that the relation $Z_R = j\,X_B$ represents a part of a more extensive bath impedance as seen in equation (17). This result is attributed to the infinite number of resonators in this model.

For a finite number of resonators, a real part behavior is observed only during a transient time following the initial excitation, and it vanishes gradually in time. As the response reaches stable conditions, only the imaginary part is retained. As the number of resonators is increased, the duration of the transient time grows, ultimately reaching an infinite value when an infinite number of resonators are used. This implies the presence of a constant real part in the impedance.

We proceed to establish an arbitrary realizable band limited formula of the resistive part of the impedance. Next, we substitute this into equation (21)a to calculate the reactive part, as follows

$$R_B(\omega) = R\frac{\omega^2}{\omega^2 + \omega_b^2}\frac{\omega_c^2}{\omega^2 + \omega_c^2}$$

$$X_B(\omega) = R\frac{\omega_c}{\omega_b + \omega_c}\frac{\omega_c\omega}{\omega^2 + \omega_b^2}\frac{\omega^2 - \omega_b\omega_c}{\omega^2 + \omega_c^2} \quad (22)$$

The resistive part equation presents a combination of High-Pass Frequency (HPF) and Low-Pass Frequency (LPF) responses. The purpose of incorporating the HPF response part, over Cataneo and Paraoanu [15] approach, is to prevent the occurrence of singularity while calculating the loop interaction inductor $L_L$ by equation (4).

By plotting the relative dimensions of the real and imaginary parts of equation (22), using arbitrary values of $\omega_b$ and $\omega_c$, we obtain the following curves

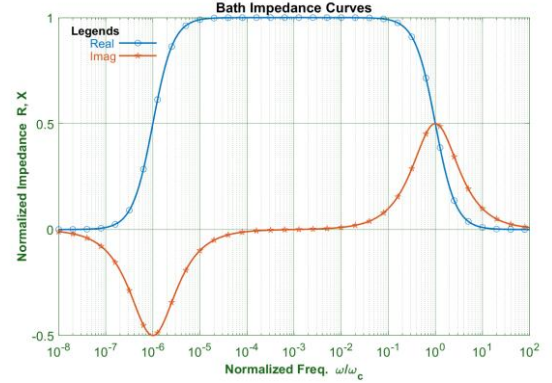

Fig. 4 • Real and imaginary parts of the bath impedance, normalized to qubit parasitic resistance $R_B$ vs. frequency normalized to LPF knee frequency $\omega_c$.

The above analysis fails to include the DC case, ω=0. In this case, the flow of current is limited only to the bath inductors, resulting in an impedance value of zero.

## 5. Thermal Noise on the Resistor

In this chapter, we will analyze the thermal properties of the resistor's model described in equation (22) and demonstrate its close relation to the well-known Johnson-Nyquist noise [24].

*Noise Spectral Density*

To calculate the spectrum density, we start with deriving the correlation function of the voltage drop across the resistor by calculating the auto-correlation function for each LC resonator at the bath electrical model, as follows

$$C_{VV}^{(k)}(t) = \langle v_k(t)v_k(0)\rangle = \langle \dot\phi_k(t)\dot\phi_k(0)\rangle_{\rho_B(\beta)} \quad (23)$$

Three functions must be calculated for (23)

$\langle \dot\phi_k(0)^2\rangle$      Flux derivative operator variance at $t=0$

$\langle \phi_k(0)\dot\phi_k(0)\rangle$      Flux-to-flux derivative covariance at $t=0$

$\dot\phi_k(t)$      Time dependent flux derivative operator

This calculation is already depicted in reference [15]. It is expressed by the integral

$$C_{VV}(t) = \int_{-\infty}^{\infty}\frac{\hbar\omega_k}{\pi}R_B(\omega_k)\mathcal{N}(\omega_k)\,e^{-i\omega_k t}\,d\omega_k \quad (24)$$

$$\mathcal{N}(\omega_k) = [n(\omega_k)+1]u(\omega_k) - n(-\omega_k)u(-\omega_k)$$

where $u(\omega)$ is the Heaviside Step-function [25], and the Bose-Einstein density function is

$$n(\omega_k) = \frac{1}{e^{\beta\hbar\omega_k}-1} \quad (25)$$

The physical interpretation of the states weighting function $\mathcal{N}(\omega)$ is attained from the limit $\lim_{T\to 0}n(\omega) = 0$, since only the positive frequency part of $\mathcal{N}(\omega)$ remains. In the limit of $T$ approaching zero, the system is only in the absorption state, with no emission. This leads to the following classification

a) $\omega > 0$   Absorption state   $\mathcal{N}(\omega) = n(\omega) + 1$
b) $\omega < 0$   Emission state   $\mathcal{N}(\omega) = n(-\omega)$ (26)

as illustrated in Fig. 5.

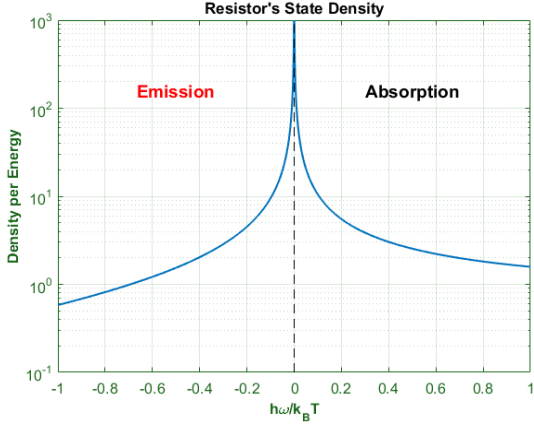

Fig. 5 • Resistor's density of states vs. normalized energy

To find the spectral density $S_{VV}(\omega)$ of the thermal noise across the resistor, we use the Weiner-Khinchin theorem [26]

$$S_{VV}(\omega) = \int_{-\infty}^{\infty} C_{VV}(t) e^{i\omega t} dt \quad (27)$$

Substituting (27) into (24), we obtain the spectral density function

$$S_{VV}(\omega) = \frac{2\hbar\omega}{1 - e^{-\beta\hbar\omega}} \cdot R \frac{\omega^2}{\omega^2 + \omega_b^2} \cdot \frac{\omega_c^2}{\omega^2 + \omega_c^2} \quad (28)$$

In order to estimate the impact of the analysis for our Josephson Junction qubit, operating at 13.5GHz, we arbitrarily set the bath LPF to 1.0THz, about 100 times higher than the qubit frequency, and the bath HPF to 1MHz, about 10,000 smaller than qubit frequency. The derived power density of the resistor bath is shown in Fig. 6

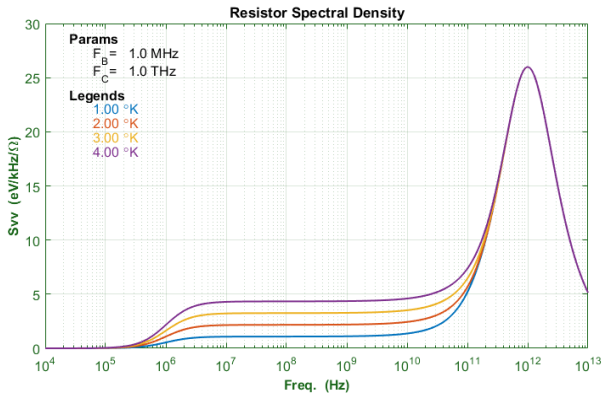

Fig. 6 • Power density of resistor's bath model

At lower frequencies, over the HPF rejection range, $\hbar\omega/k_B T \ll 1$ where the exponent expression can be approximated to the first order, the spectral density can be approximated by

$$S_{VV}(\omega) \cong 2k_B TR \quad (29)$$

which is similar to the known Johnson-Nyquist noise [24], except for the factor of 2 instead of 4 in the Johnson-Nyquist model. This difference is due to the different frequency ranges. The Johnson-Nyquist model applies for positive frequencies, while equation (29) includes both the positive and the negative ranges.

## 6. Emission And Absorption Rates

*Rotating Frame Interaction Hamiltonian*

Rewriting equation (14), we get

$$\widetilde{H}_0 = -\frac{\omega_J}{2}\widetilde{\sigma}_z + \sum_{k=1}^{\infty} \omega_{Ak} \widetilde{b}_k^\dagger \widetilde{b}_k \quad (30)$$

Since the Hamiltonian $\widetilde{H}_0$ is a diagonal matrix, the general matrix element of the Rotating Frame matrix is

$$[U_{rf}]_{i,j} = [e^{i\widetilde{H}_0 t}]_{i,j} = \delta_{ij} \cdot e^{[\widetilde{H}_0]_{i,j}} \quad (31)$$

and the rotated frame Interaction Hamiltonian is

$$\widehat{H}_I(t) = e^{i\widetilde{H}_0 t} \widetilde{H}_I e^{-i\widetilde{H}_0 t} \quad (32)$$

Substituting equation (14)d, (30) and (31) into (32), with using the general formula

$$B = U_{rf} A U_{rf}^\dagger \quad \Rightarrow \quad B_{mn} = e^{i(\omega_m - \omega_n)t} A_{mn} \quad (33)$$

we obtain the rotating frame Interaction Hamiltonian

$$\widehat{H}_I(t) = \widetilde{\sigma}_x \widetilde{\Phi}_R = \sigma_x \otimes \sum_k \omega_{Rk} \Phi_{Rk}$$

$$\sigma_x = \frac{1}{\sqrt{2}}(\sigma_+ e^{i\omega_{AJ}t} + \sigma_- e^{-i\omega_{AJ}t}) \quad (34)$$

$$\Phi_{Rk} = \frac{1}{\sqrt{2}}(\widetilde{b}_k^\dagger e^{i\omega_{kJ}t} + \widetilde{b}_k e^{-i\omega_{kJ}t})$$

From equation (13), the coupling frequency is

$$\omega_{Rk} = \frac{\sqrt{L_{AJ} L_{Ak}}}{L_L} \sqrt{\omega_{AJ} \omega_{Ak}} \quad (35)$$

*Coupling Frequency Approximation*

To calculate the coupling frequency, we first substitute equation (22) into (19) and next into (4) to obtain

$$L_L = L_J + \frac{2R}{\pi} \sum_{k=1}^{\infty} \frac{\omega_c^2}{\omega_k^2 + \omega_c^2} \frac{\Delta\omega}{\omega_k^2 + \omega_b^2} \quad ; \quad \omega_k = k\Delta\omega \quad (36)$$

Approximating the sum term by converting it to a continuous integral using the approximation $\omega_b \ll \omega_c$, we get

$$L_L = L_J + R \frac{\omega_c}{\omega_b(\omega_b + \omega_c)} \cong L_J + \frac{R}{\omega_b} \quad (37)$$

Calculation for our Josephson junction qubit renders

$$\begin{aligned} L_J &= 134pH \\ R &= 10k\Omega \end{aligned} \quad \Rightarrow \quad \omega_b \ll \frac{10k\Omega}{134pH} = 2\pi \cdot 11.9 THz \quad (38)$$

Thus, we can conclude that

$$\frac{\omega_b L_J}{R} \ll 1 \quad \Rightarrow \quad L_L \cong \frac{R}{\omega_b} \qquad (39)$$

Using this approximation in equation (9) we obtain

$$L_{AJ} \cong \left(\frac{1}{L_J} + \frac{\omega_b}{R}\right)^{-1} \cong L_J$$

$$L_{Ak} \cong \left(\frac{1}{L_k} + \frac{\omega_b}{R}\right)^{-1} \cong L_k \qquad (40)$$

Substituting this approximation into equation (13) we get

$$\omega_{AJ} \cong \omega_J \quad ; \quad \omega_{AK} \cong \omega_k \qquad (41)$$

Substituting equations (39)-(41) into equation (35), we get the coupling frequency approximation

$$\omega_{Rk} = \sqrt{J(\omega_k)\,\Delta\omega} \qquad (42)$$

where $J(\omega_k)$ is the Spectral Ohmic Density

$$J(\omega_k) = \gamma \omega_k \frac{\omega_b^2}{\omega_k^2 + \omega_b^2} \cdot \frac{\omega_c^2}{\omega_k^2 + \omega_c^2} \quad ; \quad \gamma = \frac{2}{\pi} \cdot \frac{\omega_J L_J}{R} \qquad (43)$$

Note that this expression has frequency units, unlike the ordinary spectral density which is unitless. For the contiguous limit, where $\Delta\omega \to 0$, we get

$$J(\omega) = \lim_{\Delta\omega \to 0} J(\omega_k) = \gamma \omega \frac{\omega_b^2}{\omega^2 + \omega_b^2} \cdot \frac{\omega_c^2}{\omega^2 + \omega_c^2} \qquad (44)$$

From reference [15]-par. 4.1, the emission and absorption rates are

$$\Gamma_e = \pi J(\omega_J)\left[\coth\frac{\beta\hbar\omega_J}{2} + 1\right] \qquad Ems.: Sys. \twoheadrightarrow Bath$$

$$\Gamma_a = \pi J(\omega_J)\left[\coth\frac{\beta\hbar\omega_J}{2} - 1\right] \qquad Abs.: Bath \twoheadrightarrow Sys. \qquad (45)$$

for negative and positive frequencies, respectively. Applying the relations between the frequencies: $\omega_b \ll \omega_J \ll \omega_c$, we can approximate

$$J(\omega_J) \cong \gamma \frac{\omega_b^2}{\omega_J} \quad ; \quad \gamma = \frac{2}{\pi} \cdot \frac{\omega_J L_J}{R} \qquad (46)$$

Substituting equation (46) into (45), we obtain

$$\Gamma_e = \frac{2\omega_b^2 L_J}{R} \cdot \left[\coth\frac{\beta\hbar\omega_J}{2} + 1\right] \qquad Ems.: Sys. \twoheadrightarrow Bath$$

$$\Gamma_a = \frac{2\omega_b^2 L_J}{R} \cdot \left[\coth\frac{\beta\hbar\omega_J}{2} - 1\right] \qquad Abs.: Bath \twoheadrightarrow Sys. \qquad (47)$$

$\omega_b$ will be derived in the next paragraph.

*Coherent Wave Decay Constant*

Applying the non-truncated ladder matrix for the system, in a manner similar to equation (12), using approximation (41) and changing the zero energy level for (12)b, the Hamiltonian obtained is

$$\widetilde{H} = \widetilde{H}_S + \widetilde{H}_B + \widetilde{H}_I \qquad (a)$$

$$\widetilde{H}_S = \omega_J \widetilde{\mathbf{a}}^\dagger \widetilde{\mathbf{a}} \qquad (b)$$

$$\widetilde{H}_B = \sum_{k=1}^{\infty} \omega_k \left(\widetilde{\mathbf{b}}_k^\dagger \widetilde{\mathbf{b}}_k + \frac{1}{2}\right) \qquad (c) \qquad (48)$$

$$\widetilde{H}_I = \frac{1}{2}(\widetilde{\mathbf{a}}^\dagger + \widetilde{\mathbf{a}})\sum_{k=1}^{\infty} \omega_{Rk}(\widetilde{\mathbf{b}}_k^\dagger + \widetilde{\mathbf{b}}_k) \qquad (d)$$

where $\omega_{Rk}$ is given by equation (42). Therefore, in a similar manner to (34), we get the rotated frame interaction Hamiltonian

$$\widehat{H}_I(t) = \widetilde{\Phi}_J \widetilde{\Phi}_R = \widehat{\Phi}_J \otimes \sum_k \omega_{Rk} \widehat{\Phi}_{Rk} \qquad (a)$$

$$\widehat{\Phi}_J = \frac{1}{\sqrt{2}}(\mathbf{a}^\dagger e^{i\omega_J t} + \mathbf{a} e^{-i\omega_J t}) \qquad (b) \qquad (49)$$

$$\widehat{\Phi}_{Rk} = \frac{1}{\sqrt{2}}(\widetilde{\mathbf{b}}_k^\dagger e^{i\omega_k t} + \widetilde{\mathbf{b}}_k e^{-i\omega_k t}) \qquad (c)$$

Assuming that the system's initial condition is

$$|\widehat{\psi}(0)\rangle = |\alpha_n\rangle \otimes |k\rangle$$
$$\Downarrow \qquad (50)$$
$$\widehat{\rho}(0) = |\alpha_n\rangle\langle\alpha_n| \otimes |k\rangle\langle k| = \rho_0 \otimes \rho_k$$

where $|k\rangle$ is some arbitrary bath state and $|\alpha_n\rangle$ is the coherent state of order $n$ [19]. Therefore

$$\langle \alpha_n, k|\widehat{\rho}(0)|k, \alpha_n\rangle = e^{-\alpha_n^2} \cong 0$$
$$\langle \alpha_n, k|[\widehat{H}_I(t'), \widehat{\rho}(0)]|k, \alpha_n\rangle = 0 \qquad (51)$$

Applying the perturbation theory to the Schrödinger equation, using equation (51), we get the following time dependent probability of state $|\alpha_n\rangle$, initiating from state $|\widehat{\psi}(0)\rangle$,

$$p_{\alpha k}(t) \cong \langle \alpha_n, k|\widehat{\rho}(t)|k, \alpha_n\rangle$$
$$\cong \int_0^t \int_0^t \langle \alpha_n, k|\widehat{H}_I(t')\widehat{\rho}(0)\widehat{H}_I(t'')|k, \alpha_n\rangle \, dt'dt'' \qquad (52)$$

Substitute equation (50) into (52), we get

$$p_{\alpha k}(t) \qquad (53)$$
$$\cong \int_0^t \int_0^t \langle \alpha_n, k|\widehat{H}_I(t')|k, \alpha_n\rangle\langle \alpha_n, k|\widehat{H}_I(t'')|k, \alpha_n\rangle \, dt'dt''$$
$$= \omega_{Rk}^2 \int_0^t \int_0^t F(t')G(t')G(t'')F(t'') \, dt'dt''$$

where

$$F(t) = \langle \alpha_n|\widehat{\Phi}_J(t)|\alpha_n\rangle$$
$$G(t) = \langle k|\widehat{\Phi}_{Rk}(t)|k\rangle \qquad (54)$$

From equation (49)b, assuming that $\alpha_n$ is real and positive, we get

$$F(t) = \langle \alpha_n | \widehat{\Phi}_J(t) | \alpha_n \rangle$$
$$= \frac{1}{\sqrt{2}} \langle \alpha_n | \boldsymbol{a}^\dagger e^{i\omega_J t} + \boldsymbol{a} e^{-i\omega_J t} | \alpha_n \rangle \quad (55)$$
$$= \frac{1}{\sqrt{2}} (\alpha_n^* e^{i\omega_J t} + \alpha_n e^{-i\omega_J t}) = \sqrt{2}\, \alpha_n \cos \omega_J t$$

Substituting equation (55) into (53), we derive

$$p_{\alpha k}(t) \cong \omega_{Rk}^2 \int_0^t \int_0^t G(t') G(t'') \cos \omega_J t' \cos \omega_J t''\, dt' dt'' \quad (56)$$

Calculating the sum of all the resistor's resonators, we obtain

$$p_\alpha(t) \cong 2\alpha_n^2 \omega_{Rk}^2 \cdot \int_0^t \int_0^t \sum_{k=0}^\infty \langle k | \omega_{Rk}^2 \widehat{\Phi}_{Rk}(t') \cdot p_{kk} \rho_k \cdot \widehat{\Phi}_{Rk}(t'') | k \rangle \cos \omega_J t' \cos \omega_J t''\, dt' dt'' \quad (57)$$

Following several algebraic manipulations, we get

$$\Delta p_\alpha(t) \cong 2\alpha_n^2 \int_0^t \int_0^t C_{\phi\phi}(t',t'') \cos \omega_J t' \cos \omega_J t''\, dt' dt'' \quad (58)$$
$$C_{\phi\phi}(t',t'') = \sum_{k=0}^\infty [p_{kk} \omega_{Rk}^2 \langle \widehat{\Phi}_{Rk}(t') \widehat{\Phi}_{Rk}(t'') \rangle]$$

Denoting the time difference as $\tau = t' - t''$, and due to the ergodic nature of the resistor's noise, there is no time dependence of the auto-correlation term of the last integrand. Therefore, we can write

$$s_\alpha(t) = \int_0^t \int_0^t C_{\phi\phi}(\tau) \cos \omega_J \tau\, d\tau dt' + \int_0^t \int_0^t C_{\phi\phi}(\tau) \cos \omega_J (2t' - \tau)\, d\tau dt' \quad (59)$$

where

$$C_{\phi\phi}(t) = \int_{-\infty}^\infty J(\omega_k) \mathcal{N}(\omega_k) e^{i\omega_k t}\, d\omega_k \quad (60)$$

To derive the inner integral of equation (59), and using the fact that the auto-correlation function is a time limited signal, we extend the limits of its inner integral to infinity. Using the spectral density expression of equation (27), we obtain

$$s_{\alpha 1}(t) = \frac{1}{2} \int_0^t \int_{-\infty}^\infty C_{\phi\phi}(\tau) \left( e^{i\omega_J \tau} + e^{-i\omega_J \tau} \right) d\tau dt' \quad (61)$$
$$= \frac{1}{2} [S_{\phi\phi}(\omega_J) + S_{\phi\phi}(-\omega_J)] \cdot t$$

Substitute (60) into (61), we get

$$s_{\alpha 1}(t) = -\pi J(\omega_J) \cdot t \quad (62)$$

For the second integral, in the same manner as for the first one, we extend its inner integral limits to infinity

$$s_{\alpha 2}(t) \cong \frac{1}{2} \int_0^t \int_{-\infty}^\infty C_{\phi\phi}(\tau) \left[ e^{i\omega_J(2t'-\tau)} + e^{-i\omega_J(2t'-\tau)} \right] d\tau dt' \quad (63)$$

Changing the integrand variable for the first inner sub-integral, we get

$$F_1(\omega_J, t') = \int_{-\infty}^\infty C_{\phi\phi}(2t' - t_\Delta) e^{i\omega_J t_\Delta}\, dt_\Delta \quad (64)$$

Due to the ergodic nature of the auto-correlation function, the $2t'$ term is meaningless, therefore

$$F_1(\omega_J, t') = \int_{-\infty}^\infty C_{\phi\phi}(-t_\Delta) e^{i\omega_J t_\Delta}\, dt_\Delta = S_{\phi\phi}(-\omega_J) \quad (65)$$

In a similar manner, the second inner sub-integral

$$F_2(\omega_J, t') = \int_{-\infty}^\infty C_{\phi\phi}(\tau) e^{-i\omega_J(2t'-\tau)}\, d\tau = S_{\phi\phi}(\omega_J) \quad (66)$$

Substituting (65) and (66) into (59), using (62), we get

$$s_{\alpha 2}(t) = \frac{1}{2} [S_{\phi\phi}(\omega_J) + S_{\phi\phi}(-\omega_J)] \cdot t = -\pi J(\omega_J) \cdot t \quad (67)$$

Substituting (62) and (67) into (58), we get

$$\Delta p_\alpha(t) = 2\alpha_n^2 \cdot [s_{\alpha 1}(t) + s_{\alpha 2}(t)] = -2\pi J(\omega_J) \cdot t \quad (68)$$

Therefore, using the approximation of equation (46), the initial emission rate is given by

$$\Gamma_{\alpha e} = -4\pi \alpha_n^2 J(\omega_J) \cong -8\alpha_n^2 \frac{\omega_b^2 L_J}{R} \quad (69)$$

This expression can be interpreted as

$$\Gamma_{\alpha e} \cong -8 \frac{E_0}{\hbar \omega_J} \frac{\omega_b^2 L_J}{R} \quad ; \quad E_0 \cong \hbar \omega_J \alpha_n^2 \quad (70)$$

This result holds for a coherent wave, which is a quasi-classical wave. Therefore, an emission rate of a signal with a large enough wave parameter [19] : $\alpha_n \gg 1$, can be compared to a well-known classical wave decay parameter, as depicted below.

The classical Kirchhoff junction equation for parallel RLC network is

$$0 = \frac{v}{R} + C_J \frac{dv}{dt} + \frac{1}{L_J} \int_{-\infty}^t v\, dt' \quad (71)$$

Using the Laplace transform, assuming that the initial junction voltage is $v_0$, we obtain the solution

$$v(t) = v_0 e^{-\mu \omega_J t} \left( \cos \omega_{Jd} t + \frac{\mu}{\sqrt{1-\mu^2}} \sin \omega_{Jd} t \right) \quad (72)$$
$$\omega_{Jd} = \omega_J \sqrt{1 - \mu^2}$$

Accordingly, the initial energy decaying rate is

$$\gamma_E = \frac{d}{dt}\frac{1}{2}C_J v^2\Big|_{t=0}$$
$$= \frac{1}{2}C_J v_0 \gamma_v = -\frac{1}{2RC_J}\cdot\frac{1}{2}C_J v_0^2 = -\frac{E_0}{2RC_J} \quad (73)$$

Comparing equation (73) with equation (70), we get the lower high-pass section frequency of the resistor's model

$$\omega_b = \frac{\omega_J}{4} \quad (74)$$

*Solving the System's Lindblad equation*

Substitute $\omega_b$ frequency parameter into equation (47), we obtain, following several algebraic manipulations

$$\Gamma_e = \frac{1}{8RC_J}\cdot\left(\coth\frac{\beta\hbar\omega_J}{2}+1\right) \quad Ems.:\ Sys.\twoheadrightarrow Bath$$

$$\Gamma_a = \frac{1}{8C_J}\cdot\left(\coth\frac{\beta\hbar\omega_J}{2}-1\right) \quad Abs.:\ Bath\twoheadrightarrow Sys. \quad (75)$$

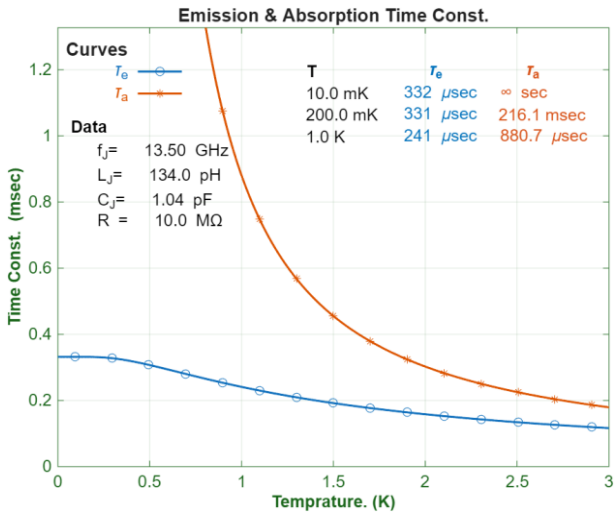

Fig. 7 • Emission and absorption time constants vs. temperature.

The dependence of these rates on temperature is demonstrated in the curves in Fig. 7. The time constants here are $T_{e,a} = 1/\Gamma_{e,a}$. As can be seen, one can use a single qubit for a reasonable time for temperatures below 1K. These emission and absorption rates, appear in the general Lindblad master equation, as follows [27]:

$$\frac{d}{dt}\rho_S = -i\,[H_S,\rho_S]$$
$$+\Gamma_e\left(\boldsymbol{a}\,\rho_S\,\boldsymbol{a}^\dagger - \frac{1}{2}\{\boldsymbol{a}^\dagger\boldsymbol{a},\rho_S\}\right) + \Gamma_a\left(\boldsymbol{a}^\dagger\rho_S\,\boldsymbol{a} - \frac{1}{2}\{\boldsymbol{a}\,\boldsymbol{a}^\dagger,\rho_S\}\right) \quad (76)$$

where

- $\rho_S = \rho_S(t)$ is the non-dissipative system's density matrix
- $H_s$ is the non-dissipative system's Hamiltonian
- $\boldsymbol{a}^\dagger, \boldsymbol{a}$ are the non-dissipative system's ladder operators
- $\{A,B\} = AB + BA$ are the anti-commutative brackets
- $\Gamma_e = \Gamma_e(t)$ is the emission rate
- $\Gamma_a = \Gamma_a(t)$ is the absorption rate

Assuming that the absorption rate is negligible relative to the emission rate, and using the expressions of equation (48), we obtain

$$\frac{d}{dt}\rho_S = -i\,\left[\frac{\omega_J}{2}\boldsymbol{\sigma_z},\rho_S\right] + \Gamma_e\left(\boldsymbol{\sigma}_-\,\rho_S\,\boldsymbol{\sigma}_+ - \frac{1}{2}\{\boldsymbol{\sigma}_+\boldsymbol{\sigma}_-,\rho_S\}\right) =$$
$$= -i\frac{\omega_J}{2}(\boldsymbol{\sigma_z}\rho_S - \rho_S\boldsymbol{\sigma_z}) + \Gamma_e\left(\boldsymbol{\sigma}_-\,\rho_S\,\boldsymbol{\sigma}_+ - \frac{1}{2}\boldsymbol{\sigma}_+\boldsymbol{\sigma}_-\rho_S - \frac{1}{2}\rho_S\,\boldsymbol{\sigma}_+\boldsymbol{\sigma}_-\right) \quad (77)$$

where

$$\widetilde{\boldsymbol{\sigma}}_+ = \frac{1}{\sqrt{2}}(\widetilde{\boldsymbol{\sigma}}_x + i\widetilde{\boldsymbol{\sigma}}_y)\ ;\ \widetilde{\boldsymbol{\sigma}}_- = \frac{1}{\sqrt{2}}(\widetilde{\boldsymbol{\sigma}}_x - i\widetilde{\boldsymbol{\sigma}}_y) \quad (78)$$

and the required density matrix

$$|\psi(t)\rangle = a|1\rangle + b|0\rangle$$
$$\rho_S = |\psi(t)\rangle\langle\psi(t)| = \begin{bmatrix} a^2 & abe^{-i\delta} \\ abe^{i\delta} & b^2 \end{bmatrix} \quad (79)$$
$$b = \sqrt{1-a^2}$$

Substituting equations (78) and (79) into (77), one gets

$$\frac{d}{dt}\rho_S = \begin{bmatrix} D & E \\ E^* & -D \end{bmatrix}$$
$$D = -\Gamma\,a^2\ ;\ E = -\left(\frac{\Gamma}{2}+i\omega_J\right)a\sqrt{1-a^2}\,e^{-i\delta} \quad (80)$$

This is a system of two equations with two variables, $a(t), \delta(t)$, Assuming a constant decoherence function $\Gamma$, as follows

$$\frac{d}{dt}(a^2) = -\Gamma\,a^2$$
$$\frac{d}{dt}\left(a\sqrt{1-a^2}\,e^{-i\delta}\right) = -\left(\frac{\Gamma}{2}+i\omega_J\right)a\sqrt{1-a^2}\,e^{-i\delta} \quad (81)$$

The solution of the two equations with two variables, $a(t), \delta(t)$ is given by

$$a = a_0 e^{-\frac{\Gamma}{2}t}\ ;\ b = \sqrt{1-a_0^2}\,e^{i\delta}$$
$$\delta = -\sqrt{\delta_0^2 + 2i\left(a_0\sqrt{1-a_0^2} - a\sqrt{1-a^2}\right) - i(\Gamma+i2\omega_J)t} \quad (82)$$

where the negative sign of $\delta$ is explained in the following paragraph.

For a sufficiently long time and $\Gamma \ll \omega_J$ one can approximate

$$\delta \cong -\sqrt{-i(\Gamma+i2\omega_J)t} \cong -\sqrt{2\omega_J}\left(1 - i\frac{\Gamma}{4\omega_J}\right)\sqrt{t} \quad (83)$$

therefore

$$e^{i\delta} \cong e^{-\frac{\Gamma}{2\sqrt{2\omega_J}}\sqrt{t}}\cdot e^{i\sqrt{2\omega_J t}} \quad (84)$$

which has an absolute value that decays over time, as expected from a physical system.

To demonstrate a numerical behavior of equations (82), we chose typical parameters in Fig. 8. In this example, the resonance frequency of the qubit is 13.5GHz. The qubit operates at a temperature of 10mK, with an emission time constant of 332μsec as designated in Fig. 7 above. The initial state, $|+\rangle = \frac{1}{\sqrt{2}}(|0\rangle + |1\rangle)$, is with equal probabilities for the two first states of the qubit and zero phase angle $\delta_0$ between the states.

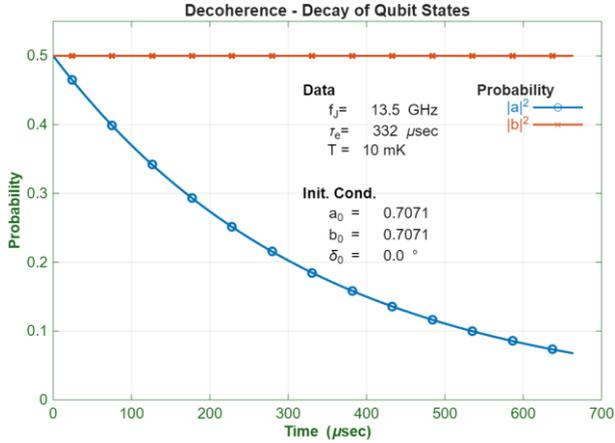

Fig. 8 • State probability vs. time showing decoherence – state $|1\rangle$ decay. $a_0 = 1/\sqrt{2}$, $\delta_0 = 0$.

Figure 9 shows the decoherence process for a longer time scale, to emphasize the slow decay of $|0\rangle$ state. As expected, state $|1\rangle$ decays much faster than state $|0\rangle$.

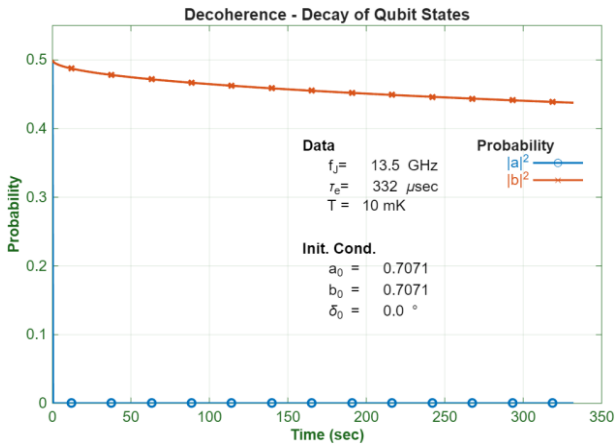

Fig. 9 • Decoherence Example for $a_0 = 1/\sqrt{2}$, $\delta_0 = 0$. State probability vs. Time. Long time scale to emphasize state $|0\rangle$ decay

## 7. CONCLUSIONS

The model parameters are obtained by comparing coherent wave time response to classical circuit response. The numerical results based on extensive analysis confirm the experimental results that over a temperature of about 0.6K, a high degradation of the coherence is observed. The decaying rate is dependent on the state, the $|1\rangle$ state decays much faster than the $|0\rangle$ state.

The calculation presented in this work suggests a process that opens a way for a deterministic method to analyze the qubit decoherence for a variety of internal parasitic resistance types.